\documentclass{aims}
\usepackage{amsmath}
 \usepackage[colorlinks=true]{hyperref}
\hypersetup{urlcolor=blue, citecolor=red}
\def\currentvolume{X}
 \def\currentissue{X}
  \def\currentyear{200X}
   \def\currentmonth{XX}
    \def\ppages{X--XX}

\usepackage{graphicx}
\usepackage{bm}

\newcommand{\remove}[1]{} 
\usepackage{url}
\usepackage{algorithmic}
\usepackage{algorithm}
\usepackage{ amsmath}
\usepackage{dcolumn}

\newcommand{\cent}{\mathrm{cr}}
\newcommand{\ncent}{\mathrm{ncr}}

\newcommand{\pr}{\mathrm{pr}}

\newcommand{\C}{x_c}
\newcommand{\N}{x_n}
\newcommand{\WC}{\mathrm{\mathcal{T}}}
\newcommand{\WN}{\mathrm{\mathcal{R}}}

\title[Rethinking Centrality]
      {Rethinking Centrality: The Role of Dynamical Processes in Social Network Analysis}

\author[Rumi Ghosh and Kristina Lerman]{}


 \email{rumig@usc.edu}
 \email{lerman@isi.edu}

\begin{document}
\maketitle

\centerline{\scshape Rumi Ghosh }
\medskip
{\footnotesize
 \centerline{HP Labs}
   \centerline{1501 Page Mill Road}
   \centerline{Palo Alto, CA 94304, USA}
} 

\medskip

\centerline{\scshape Kristina Lerman}
\medskip
{\footnotesize
 \centerline{USC Information Sciences Institute}
   \centerline{4676 Admiralty Way}
   \centerline{Marina del Rey, CA 90292, USA}
}

\bigskip


\begin{abstract}
Many popular measures used in social network analysis, including centrality, are based on the random walk. The random walk is a model of a stochastic process where a node interacts with one other node at a time. However, the random walk may not be appropriate for modeling social phenomena, including epidemics and information diffusion, in which one node may interact with many others at the same time, for example, by broadcasting the virus or information to its neighbors. To produce meaningful results, social network analysis algorithms have to take into account the nature of interactions between the nodes.
In this paper we classify dynamical processes as conservative and non-conservative and relate them to well-known measures of centrality used in network analysis: PageRank and Alpha-Centrality. We demonstrate, by ranking users in online social networks used for broadcasting information, that non-conservative Alpha-Centrality generally leads to a better agreement with an empirical ranking scheme than the conservative PageRank.
\end{abstract}

\section{Introduction}
Social network analysis algorithms examine topology of a network in order to find interesting structure within it. It has been recognized recently, however, that network structure is the product of both its \emph{links} and the \emph{dynamical processes} taking place on the network, which determine how ideas, pathogens, or influence flow along social links~\cite{Borgatti05,Borgatti06,Lambiotte10}.
Borgatti~\cite{Borgatti05,Borgatti06}, for example, argued that a node's centrality, a measure often used to identify important or influential actors in a social network, gives a summary of its participation in the flow taking place on the network. An appropriate centrality for a given network, therefore, is one whose assumptions match the details of the flow.
Some of the best-known measures of centrality, such as  PageRank~\cite{PageRank} and its variants~\cite{Jeh03}, are based on random walk-like phenomena~\cite{Tong06,Fortunato07}.
A random walk on a graph is a stochastic process that starts at some node, and at each time step transitions to a randomly selected neighbor of the current node. Variants of the random walk are often used to model flows in physical systems, e.g., chemical and heat diffusion, and can be used to model social phenomena resulting from one-to-one interactions, such as Web surfing or phone conversations. Random walks, however, do not model many phenomena of interest to social scientists, such as adoption of innovation~\cite{Rogers03,Bettencourt05}, spread of epidemics~\cite{Anderson91,Hethcote00} and word-of-mouth recommendations~\cite{Goldenberg01}, viral marketing campaigns~\cite{Kempe03,Iribarren09}, growth of social movements~\cite{Centola07} and information diffusion~\cite{Lerman10icwsm}. These phenomena are usually modeled as an epidemic process, where rather than choosing \emph{one} neighbor, an activated or ``infected'' node will attempt to activate \emph{all} its neighbors. For example, on the social media site Twitter users broadcast their posts, called tweets, to all their followers. Similarly, in an epidemic, an infectious person will pass the virus to all susceptible contacts.
Therefore, unlike the random walk, which conserves the amount of the diffusing substance, epidemic processes are fundamentally \emph{non-conservative}.

This paper makes two contributions. First, we classify dynamical processes as conservative and non-conservative and study their relationship to two well-known centrality measures: PageRank and Alpha-Centrality. PageRank~\cite{PageRank}, originally used in Google's successful search engine, gives the steady state distribution of a conservative dynamical process (specifically, random walk with random restarts~\cite{Tong06,Fortunato07}). Alpha-Centrality~\cite{Bonacich87} 
measures the number of paths of any length between two nodes, exponentially attenuated by parameter $\alpha$, so that longer paths contribute less to centrality than shorter paths. We demonstrate that Alpha-Centrality gives the steady state distribution of a class of non-conservative dynamical processes while $\alpha$ is bounded by inverse of the largest eigenvalue of the adjacency matrix of the graph. This quantity, called the epidemic threshold, governs the behavior of many non-conservative processes in networks, for example, the spread of a virus along social links~\cite{Pastor-Satorras2001,Wang03}. When the effective transmissibility of the virus is below this threshold, it will die out~\cite{Wang03,Prakash11}, but above the threshold it will reach a finite fraction of all nodes, resulting in an epidemic. Our analysis provides an intuitive explanation for the location of the epidemic threshold, and a further demonstration of the fundamental connection between network structure and dynamics.

The second contribution of the paper is an empirical study of the ability of PageRank and Alpha-Centrality to identify influential social media users.
Specifically, we study the online social networks of the social news aggregator Digg and the microblogging service Twitter, both of which are used by people to share news stories and other content with their followers. The spread of information is often modeled as an epidemic process~\cite{Watts07,Ghosh11wsdm,Versteeg11,Goel12}, hence it has a non-conservative flavor. We define two empirical measures of influence based on user activity, and rank users according to these measures. We show that non-conservative Alpha-Centrality generally leads to a better agreement with the activity-based rankings than conservative PageRank.
While the effect of dynamical processes on centrality was studied theoretically and in simulation~\cite{Borgatti05}, our work provides an empirical demonstration that the choice of centrality impacts our ability to identify important people in real-world social networks.

This paper is organized as follows. In Section~\ref{sec:classes} we provide a description of conservative and non-conservative dynamical processes and demonstrate, in Section~\ref{sec:structure}, that Alpha-Centrality gives the steady state distribution of a non-conservative dynamical process, for example, a spreading epidemic. Then, in Section~\ref{sec:OSN}, we compare Alpha-Centrality to PageRank on the task identifying influential social media users and show that Alpha-Centrality gives a better agreement with empirical measures of influence. We conclude with a summary of related work and a conclusion.

\remove{
Our specific contributions are:
\begin{itemize}
\item Mathematically describe conservative and non-conservative dynamical processes (Section~\ref{sec:classes})
\item Demonstrate that Alpha-Centrality gives the steady state distribution of a non-conservative dynamical process (Section~\ref{sec:structure})
\item Evaluate how well Alpha-Centrality and PageRank predict influential social media users, and show that Alpha-Centrality gives a better agreement with an empirical measure of influence (Section~\ref{sec:OSN})
\end{itemize}
}

\section{Classification of Dynamical Processes}
\label{sec:classes}
We represent a network by a directed  graph $G = (V,E)$ with $V$ nodes and $E$ edges. The adjacency matrix of the graph is defined as: $A[u,v]= 1$ if $(u,v) \in E$; otherwise, $A[u,v]= 0$. Also, $A[u,u]=0$, The set of  out-neighbors of $u$ is $\lbrace v \in V \vert (u,v) \in E \rbrace$; and the set of in-neighbors is $\lbrace v \in V \vert (v,u) \in E \rbrace$. Another important quantity is the diagonal out-degree matrix $D$, which is defined as $D[i,i]=\sum_j A[i,j]=Ae^T$ and $D[i,j]=0$ $\forall$ $i\neq j$. Here, $e$ is a $|V|$-dimensional row vector of ones, and $e^T$ is its transpose.

A dynamical process is mediated by interactions between nodes, which can be thought to distribute some quantity, or weight, on a network.
Let the $|V|$-dimensional vector $x$ represent the weight of each node  at time $t$. A dynamical process is described mathematically by a function $F_t(x)$ that maps the weight vector at time $t$ to the weight vector at time $t+1$.

\subsection{Conservative Processes}
A stochastic process 
is \emph{conservative} if it simply redistributes the weights among the nodes of the graph, with the total weight remaining constant:
$||x||_1 = ||F^c_t(x)||_1$, where $||.||_1$ represents the $L_1$-norm of the argument, i.e., $||x||_1=\sum_i |x_i|$. 

To give an intuition for the mathematical formulation of conservative processes,  imagine a  society where nodes interact by redistributing money among themselves, and the money cannot be created or destroyed. Let $\C(t)$ be the amount of money each node has, and $\Delta(t)$  the amount it {receives}, at time $t$. Suppose that at each time step a node retains a fraction $(1-\alpha)$ of the amount it received in the previous step and redistributes the rest among its neighbors.  Let \emph{transfer matrix} $\WC[p,q]$ represent the fraction of the amount transferred by node $p$ to $q$. Therefore, the amount of money nodes receive  at time $t+1$ is
$\Delta(t+1)=\alpha \Delta(t)\WC$.
The transfer matrix encodes the rules of interaction.
If each member  \emph{divides} $\alpha \Delta(t)$ equally amongst her out-neighbors, then $\WC=D^{-1}A$.

Step by step, conservative process looks as follows. Initially, the amount each node receives is $\Delta(0)=\C(0)$.   At  time $t=1$ each node keeps  $(1-\alpha)$ of that amount and divides the rest among its out-neighbors, who receive  $\Delta(1) =\alpha \Delta(0)\WC=\alpha \C(0)\WC$. At time $t=2$, each node retains $(1-\alpha)$ of the amount it received from in-neighbors at  $t=1$, and divides the rest among its out-neighbors, who receive
$\Delta(2)= \alpha \Delta(1)\WC=\alpha^2 \C(0)\WC^2$, and so on.
The total weight (or amount of money) the nodes have  at time $t$, $\C(t)$, is the amount they retained from all previous time steps and the amount they received from in-neighbors at time $t$:
\begin{eqnarray}
\C(t) &= & (1-\alpha)\sum_{k=0}^{t-1}\Delta(k) + \Delta(t)   \nonumber \\
& =& \sum_{k=0}^{t-1}(1-\alpha) \alpha^{k} \C(0) \WC^{k}+\alpha^{t} \C(0) \WC^{t} \nonumber \\
 &= &(1-\alpha)\C(0)+\alpha \C(t-1)\WC.
                  \label{eq:Cons8}
\end{eqnarray}

\noindent As $t\to\infty$, this equation reduces to
\begin{eqnarray}
\C(t\to\infty) &=&(1-\alpha)\C(0)+\alpha \C(t\to\infty)\WC \nonumber\\
&=&(1-\alpha)\C(0){(I-\alpha \WC)}^{-1}
  \label{eq:Cons10}
\end{eqnarray}

The transfer matrix $\WC$ is a stochastic matrix, since its rows sum up to 1.
If, instead of distributing evenly among neighbors, each node decided to keep a portion $\delta$ for itself, this variant of a conservative process would be governed by the transfer matrix:
\begin{equation}
\WC= \delta I+(1-\delta) D^{-1} A.
\label{eq:Wcons}
\end{equation}

Random walk on a graph is a prototypical conservative process, since the probability to find a walker on any node of the graph is always one. There exist many flavors of random walk. One of them is the widely studied random walk with random restarts~\cite{PageRank,Boldi05pagerankas,Fortunato07}, which can be described mathematically as follows. Let the initial probability to find  the walker on any node be uniform, i.e., $\C(0)=e\frac{1}{|V|}$. At  any time, with probability $\alpha$ the  walker at node $p$ randomly chooses one of the out-neighbors of $p$ and jumps to it. With probability $(1-\alpha)$, it randomly chooses any node on the graph and jumps to it. Let matrix $S$ encode the probability of jumping to any node, $S[p,q]=\frac{1}{|V|}$, and $\WC=D^{-1}A$.
Then the probability of finding the random walker at node $q$ at time $t$ is given by
\begin{eqnarray}
\C(t) & = & (1-\alpha)\C(t-1)S+\alpha \C(t-1)\WC \nonumber \\
& = & (1-\alpha)\C(0)+\alpha \C(t-1)\WC, \nonumber
\end{eqnarray}
\noindent which  is exactly the same as Eq.~\ref{eq:Cons8}.

\subsection{Non-Conservative Processes}
A  stochastic process where the total weight  can change over time is \emph{non-conservative}: $||x||_1 \ne ||F^n_t(x)||_1$.
To illustrate the difference between conservative and non-conservative processes, we return to our hypothetical society. Again, imagine that each node  has some amount of money, however, it also has a money minting machine, so that instead of dividing the money it receives among its out-neighbors, it can give each neighbor the same amount by printing extra as needed.

Let $\Delta(t)$ represent the amount of money each node receives at time $t$. At the next time step, each node gives a fraction $\alpha$ of this amount to each of its out-neighbors, printing extra as needed. The  additional amount it produces  can be expressed using the \emph{replication matrix} ${\WN}=A$. Therefore, $\Delta(t+1)=\alpha \Delta(t)\WN$.
Initially, let $\Delta(0)=\N(0)$. At time $t=1$, each node prints $\alpha  \Delta(0)$ for each out-neighbor:
$\Delta(1)=\alpha  \Delta(0) \WN  = \alpha \N(0) \WN$.
Continuing this process,  additional amount out-neighbors receive at time $t$ is $\Delta(t) =  \alpha \Delta(t-1) \WN= \alpha^t \N(0) \WN^t$.
The total amount each node has at time $t$ is obtained by summing what it received from in-neighbors at previous time steps:
\begin{eqnarray}
\N(t)&=& \sum^{t}_{k=0} \Delta(k)=\sum_{k=0}^{t} \N(0) (\alpha\WN)^k \nonumber \\
 &=&\N(0)+\alpha\N(t-1)  \WN
\label{eq:NonCons2}
\end{eqnarray}
\noindent
 At time $t\to\infty$, Eq.~\ref{eq:NonCons2} reduces to
\begin{equation}
\N(t\to\infty) = \N(0){\sum_{k=0}^{t\to\infty}  (\alpha \WN)^k},
 \label{eq:NonCons4}
\end{equation}
\noindent which can be solved to yield
\begin{eqnarray}
\N(t\to\infty) &=& \N(0)+ \N(t\to\infty)(\alpha \WN) \nonumber \\
&=&{\N(0)}{(I-\alpha \WN)}^{-1}.
 \label{eq:NonCons5}
\end{eqnarray}
This expression is defined for  ${\alpha}<1/{\lambda_1}$, where $\lambda_1$ is the largest eigenvalue  of $\WN$.

More generally, if along with producing $\alpha$ of what it receives from each in-neighbor, a node also produces a portion $\delta$ of this amount for itself, this leads to a more general form of the replication  matrix: 
\begin{equation}
\label{eq:Wnoncons}
\WN= \frac{\delta}{\alpha} I+A.
\end{equation}

\subsubsection{Non-Conservative Dynamics and Epidemic Threshold}
\label{sec:threshold}
Non-conservative processes provide a useful framework for thinking about epidemics and other contact processes and lead to insights into the relation between dynamical processes and network structure.
Consider a virus spreading on a network, where at each time step, a contagious node may infect its susceptible neighbors with probability $\mu$ (virus birth rate). At each time step, an infected node may also be cured with probability $\beta$ (virus curing rate). Wang et al.~\cite{Wang03} modified existing models of SIS dynamics~\cite{Bailey:1975} for use on networks. The probability $p_{i,t}$ that  node $i$ is infected at time $t$  can be written in matrix notation as~\cite{Wang03}:
\begin{equation}
\label{eq:epidemic}
P_t=P_{t-1}\big((1-\beta) I+\mu A\big) =P_0 \big((1-\beta)I+\mu A \big)^t ,
\end{equation}
\noindent where $P_t$ is a vector $(p_{1,t},\ p_{2,t},\ \ldots)$, and $P_0$ is the initial probability of infection.\footnote{This model holds true only when $p_{i,t}$ is very small and there may be situations where $p_{i,t}>1$. Therefore a more accurate interpretation is that the probability of infection is proportional to $p_{i,t}$.}
$P_t$ is exactly equal to the additional weight, $\Delta(t)$, accrued  by a non-conservative process  in Eq.~\ref{eq:NonCons2}, with  $\WN= \frac{1-\beta}{\mu} I+A$ and $\alpha=\mu$. Therefore, a SIS-type epidemic is an example of a non-conservative dynamic process.

In the model in Eq.~\ref{eq:epidemic}, there exists a threshold $\mu_c$ such that when the effective transmissibility of virus $\mu/\beta < \mu_c$, it will die out, and for $\mu/\beta >\mu_c$ it will spread to a significant portion of the network. For any network, regardless of the details of the spreading mechanism~\cite{Prakash11}, this threshold is given by   the inverse of the largest eigenvalue of the adjacency matrix $A$, $\mu_c=1/|\lambda_1|$~\cite{Wang03}, what is known as the spectral radius of the graph.
In numerical experiments we simulated epidemics on different graphs using the independent cascade model~\cite{Versteeg11}. We found that the observed threshold where epidemics began to reach many nodes was consistent with the spectral radius of the respective graph.

Threshold behavior appears to be a generic property of non-conservative dynamics.
As shown in the Appendix, the expected path length of a non-conservative process, i.e., how far the process spreads as $t\to \infty$, is $l=({1-\alpha\lambda_1})^{-1}$ for $\alpha<{1}/{|\lambda_1| }$ and $l \sim O(t)$ for $\alpha > {1}/{|\lambda_1|}$. Therefore, expected path length $l$  diverges as $\alpha$ approaches ${1}/{|\lambda_1|}$ from below. This is a hallmark of critical behavior. For non-conservative processes, the critical behavior is associated with the epidemic threshold, below which the non-conservative process reaches very few nodes, but above which is reaches a significant fraction of all nodes.

There is another way to think about thresholds. Among epidemiologists, the principal quantity of interest is the reproductive number, $R_0$~\cite{Dietz93}. Intuitively, this quantity is just average number of new infections caused by a single infected person. If $R_0 > 1$, each infection creates new infections indefinitely, and results in an epidemic, while for $R_0<1$, the disease eventually dies out. Naively, the reproductive number should just be the average degree times the transmissibility, or contagiousness of the virus.  For the Digg follower graph, for example, the average degree $\langle k \rangle\approx 6$ so $R_0 \approx 6 \lambda$, where $\lambda$ is the transmissibility of the virus. In that case, an epidemic threshold at $R_0=1 \rightarrow \lambda_c \approx 1/6$, much higher than we observed in simulations of an SIR epidemic (using independent cascade model) on the Digg follower graph~\cite{Versteeg11}. While heterogeneous degree distribution (a common property of social networks) can lower the threshold compared to this prediction~\cite{VespignaniBook}, this computation is not simple, making the basic reproductive number less useful in characterizing epidemics in social networks.

\section{Dynamical Processes and Centrality}
\label{sec:structure}
The complex interplay between network structure and dynamics has broad implications for social network analysis.
Take the task of identifying  influential or prestigious actors in a social network. Over the years many different centrality measures have been developed to address this task, including degree centrality, betweenness centrality~\cite{Freeman79}, eigenvector centrality~\cite{bonacich72factoring}, PageRank~\cite{PageRank} and Alpha-Centrality~\cite{Bonacich87}, among many others.
Applied to the same network, however,  each measure leads to a different, even conflicting notion, of who the central actors are. In order to make sense of the scores produced by each centrality measure, it is important to consider the nature of the dynamical process on the network.

\subsection{Centrality Measures}
We study PageRank and Alpha-Centrality, two widely used measures of centrality, and show their relationship to conservative and non-conservative processes.

\paragraph{PageRank}
A PageRank vector ${\pr}_{\alpha}(s,t)$ is the steady state probability distribution of a random walk with restarts with a damping factor $\alpha$ (restart probability= $1-\alpha$).  The starting vector $s$, gives the probability distribution for where the walk transitions to after restarting.  The transfer matrix encodes the transition probabilities of a random walk on the network, $W = D^{-1}A$. PageRank vector ${\pr}_{\alpha}(s)$ is the unique solution of:
\begin{equation}
\label{eq:pr1}
{\pr}_{\alpha}(s) = (1-\alpha) s +  \alpha {\pr}_{\alpha}(s)W
\end{equation}

Equation~\ref{eq:pr1} is identical to the steady state solution of the linear conservative dynamic process given by Eq.~\ref{eq:Cons10}  
where $W=\WC=D^{-1}A$ and $s=\C(0)$. Therefore, \emph{PageRank is the steady state solution of a conservative process}, and it is a {conservative measure}.
Other measures derived from the random walk, such as betweenness centrality, are also conservative.

\paragraph{Alpha-Centrality}
Alpha-Centrality measures the total number of paths from a node, exponentially attenuated by their length. Bonacich introduced this measure~\cite{Bonacich87} as a generalization of the index of status proposed by Katz~\cite{Katz53}, and it is sometimes referred to as Bonacich centrality. It is also similar to the communicability index recently explored by the physics community~\cite{Estrada12}.
For an attenuation parameter $\alpha$, Alpha-Centrality vector $\cent_{\alpha}(s)$ is the solution of:
\begin{equation}
\cent_{\alpha}(s)=s+ \alpha \cent_{\alpha}(s)A,
\label{a-cen1}
\end{equation}
\noindent where the starting vector $s$ is taken as indegree centrality, $s=eA$~\cite{Bonacich01}, with $e$ a row vector of ones. Equation~\ref{a-cen1} holds while $|\alpha| < {1}/{|\lambda_{1}|}$,  the spectral radius of the network.  This bound, in fact, is the same as the epidemic threshold (Section~\ref{sec:threshold}). For positive values, parameter $\alpha$  determines how far, on average, a node's effect will be felt and sets the length scale of interactions.\footnote{Bonacich proposed to use $\alpha<0$ case to model power relations in social networks. Our focus here is on quantifying influence; therefore, we study $\alpha \ge 0$ case. } When $\alpha$ is small, Alpha-Centrality probes only the local structure of the network. As $\alpha$ grows, more distant nodes contribute to the centrality score of a given node~\cite{Ghosh11physrev}. As $\alpha \to 1/{\lambda_1}$,  the length scale of interactions diverges (Sec.~\ref{sec:threshold}) and it becomes a global measure.

One difficulty in using Alpha-Centrality is that it is not defined for $\alpha \ge 1/\lambda_1$. We recently introduced \emph{normalized Alpha-Centrality} that overcomes this problem~\cite{Ghosh11physrev}. It normalizes the score of each node by the sum of the Alpha-Centrality scores of all the nodes. The new measure avoids the problem of bounded parameters while retaining the desirable characteristics of Alpha-Centrality, namely its ability to differentiate between local and global structures.
Normalized Alpha-Centrality $\ncent_{\alpha}(s)$ is written as:
\begin{equation}
\ncent_{\alpha}(s)  = \frac{1}{||\cent_{\alpha}(s)||_1} \cent_{\alpha}(s)
\label{n-a-cen-summation}
\end{equation}
\noindent This is defined for $0 \le \alpha \le 1\ (\alpha \neq {1}/{|\lambda_1|})$.
This value changes with $\alpha$ for $\alpha < {1}/{\lambda_1}$.
For $\alpha > {1}/{\lambda_1}$, normalized Alpha-Centrality is independent of $\alpha$ and the ordering found by normalized Alpha-Centrality in this parameter range is equivalent to the ordering found by eigenvector centrality \cite{Ghosh10snakdd}.

Alpha-Centrality and its normalized version are  equivalent to Eq.~\ref{eq:NonCons4},  with the initial distribution of weight given by $\N(0)=c\cdot s$, where $c=1$ for Alpha-Centrality  and
$$c= \frac{1}{\sum_{i,j}\sum_{k=0}^{t \to \infty} \alpha^k A^k[i,j]}=\frac{1}{||(I- \alpha A)^{-1}||_1}$$
for normalized Alpha-Centrality. Note that we use notation $||M||_1= \sum_{i,j} M[i,j]$ for any matrix $M$.
Therefore, \emph{(normalized) Alpha-Centrality is the steady state solution of a non-conservative dynamic process}. Variations of non-conservative dynamics lead to other non-conservative measures of centrality, such as degree centrality, Katz index~\cite{Katz53}, SenderRank~\cite{Kiss08}, and eigenvector centrality~\cite{bonacich72factoring}.

\subsection{Choosing Appropriate Centrality Measure}
When applied to the same network, different measures of centrality may lead to different, often incompatible, views of who the central actors are.
The natural question to ask is: Which centrality measure is appropriate for a given network? The choice of centrality must be motivated by details of the dynamical process taking place on the network~\cite{Borgatti05}. Thus, a conservative measure such as PageRank is appropriate for analyzing networks on which conservative processes are taking place, for example, web surfing or money  exchange. However, for a social network on which information or epidemics are spreading, the non-conservative  Alpha-Centrality may be more appropriate.

\section{An Empirical Study of Centrality}
\label{sec:OSN}
In this section we use social media data to evaluate the claim that \emph{the measure that best identifies central nodes is one that captures details of the dynamical process taking place on the network}. Social media sites such as Facebook, Twitter, and Digg have become important hubs of social activity and conduits of information. Correctly identifying central or influential users in these networks can have far-reaching consequences for identifying noteworthy content, targeted information dissemination, and other applications. While a variety of methods~\cite{Cha10icwsm,Lee10www,Romero10,Bakshy11} have been used to identify influential social media users, each measure produces different results, with no clear understanding of when it is appropriate.
Fortunately, by exposing user activity, social media provides a rare opportunity to study the role of dynamic processes on networks.

Both Digg and Twitter allow users to create social networks by listing others as friends. The friend relationship is asymmetric. When user $A$ lists $B$ as a friend ($A \to B$), $A$ follows $B$'s activity, but not vice versa. We call $A$ the follower of $B$ (or fan on Digg). When follower graph is represented in matrix form, a user's indegree measures the number of followers she has, and her outdegree the number of friends she follows.

By submitting a story to Digg (or tweeting a URL to a story on Twitter), a user broadcasts it to her followers. When another user  votes for the story, she re-broadcasts it to her own followers.  Broadcast-driven information diffusion has a non-conservative flavor; therefore, a non-conservative centrality measure should better identify influential users.

We analyzed {information diffusion} on the follower graphs of  Digg and Twitter and used this data to construct an empirical estimate of user influence. We then compared how different centrality measures compared to the empirical measure of influence.

\subsection{Data Sets}
\label{sec:data}
The Digg dataset\footnote{http://www.isi.edu/$\sim$lerman/downloads/digg2009.html} contains more than 3 million votes on some 3500 stories promoted to Digg's front page in June 2009. More than 139K distinct users voted for at least one story in the data set (submission counts as the story's first vote). We call these users \emph{active} users. Next, we extracted the friendship links created by {active} users and constructed a follower graph that contained active users who were following the activities of others.
However, only about 71K active users listed others as friends, resulting in network with around 300K users and over 1 million links.

The Twitter data set was collected over the period of three weeks in October 2010 using the Gardenhose streaming API. We focused on tweets that included a URL in the body of the message, usually shortened by some URL shortening service, such as bit.ly or tinyurl. In order to ensure that we had the complete retweeting history of each URL, we used Twitter's search API to retrieve all tweets containing that URL. Users who tweeted the URL are considered \emph{active}. Data collection process resulted in more than 3 million posts tweeted by 816K users which mentioned 70K distinct shortened URLs. Next, we used the REST API to collect followers of  each active user, keeping only those followers who themselves were active, i.e., tweeted at least one URL during data collection period. The resulting follower graph had almost 700K nodes and over 36 million edges. While filtering out non-active followers will change results of centrality calculations, we argue that this is an appropriate simplification to make, both conceptually and to keep the graph of a computationally manageable size. We argue that inactive users do not contribute to information spread, and should not be considered in calculations of centrality.

While voting on Digg represents pure information diffusion (in contrast to Twitter, Digg user can vote only once for a story), tweeting activity in our sample encompassed diverse behaviors from pure information diffusion of newsworthy content to orchestrated manipulation campaigns,  robo-tweeting, advertising and spam. Since our analysis applies only to information diffusion-type behavior, we have to filter out latter activities. We used a method described in \cite{Ghosh11snakdd} to automatically classify tweeting behaviors using two  information theoretic features. The first feature is the entropy of the distribution of distinct users who re-tweeted the URL. The second feature is the entropy of the distribution of time intervals between successive re-tweets of the same URL. We showed that these two features alone were able to accurately separate re-tweeting activity into meaningful classes. High user entropy implies that many different people re-tweeted the URL, with most people re-tweeting it once. High time interval entropy implies presence of many different time scales, which is a characteristic of human activity. In contrast, low time interval entropy implies that URL is retweeted at one or few regular time intervals, which is characteristic of automated (possibly spam) activity. In this paper, we focus  on those URLs from the data set which are characterized by high ($>3$) user and time interval entropies. These parameter values are associated with the spread of news-worthy content and excludes robotic spamming and manipulation campaigns driven by few individuals.

\subsection{Empirical Estimates of Influence}
Katz and Lazarsfeld~\cite{Katz55} defined influentials as ``individuals who were likely to influence other persons in their immediate environment.'' In the years that followed, many attempts were made to identify people who influenced others to adopt a new practice or product~\cite{Brown87}.
The rise of online social networks has allowed researchers to trace the flow of information through social links on a massive scale. Using the new empirical foundation, some researchers proposed to measure a person's influence in social media locally, by the number of votes or retweets from followers her posts generate~\cite{Ghosh10snakdd,Bakshy11}, or globally, by the size of cascades her posts trigger~\cite{Kempe03,Bakshy11}.
Alternatively, Trusov \emph{et al.}~\cite{Trusov10} defined influential people in an online social network as those whose activity stimulates those connected to them to increase their activity, while Cha \emph{et al.}~\cite{Cha10icwsm} used the total number of retweets and mentions, including from people not connected either directly or indirectly to the submitter, to measure user influence on Twitter.

Following these works, we measure influence by analyzing user activity in social media. Suppose that a user posts new information on Digg or Twitter, specifically, a URL to a news story. We refer to this user as the story's \emph{submitter}. Whether or not her follower will re-broadcast the story (i.e., retweet it on Twitter or vote for it on Digg)  depends on its \emph{quality} and  \emph{submitter's influence}. We assume that story's quality is uncorrelated with the submitter.\footnote{This may seem like a strong statement, but as other studies of Digg show~\protect\cite{Hogg09icwsm,Hogg12epj}, how interesting a story is to submitter's followers does not depend on who the submitter is, at least not on Digg.} Therefore, we can average out its effect by aggregating over all stories submitted by the same user. We claim that the residual difference between submitters can be attributed to variations in influence. We use two empirical measures of submitter's influence: ($i$) the average number of times her submissions are re-broadcast by her followers (local influence~\cite{Bakshy11}), and ($ii$) average size of the cascades her posts trigger (global influence~\cite{Bakshy11}).

%
\begin{figure}[tbh]
\begin{tabular}{@{}c@{}c@{}}
\multicolumn{2}{c}{Digg} \\
  \includegraphics[width=0.5\linewidth]{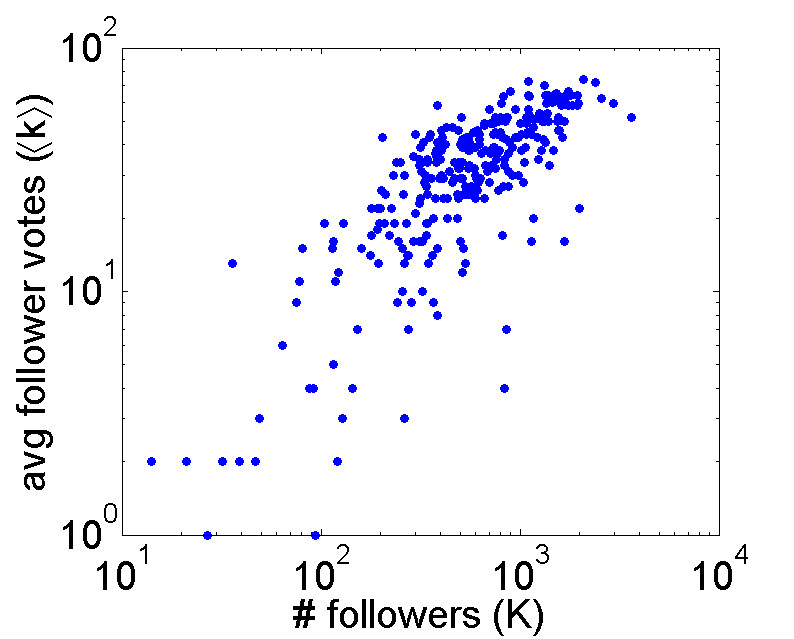} &
   \includegraphics[width=0.5\linewidth]{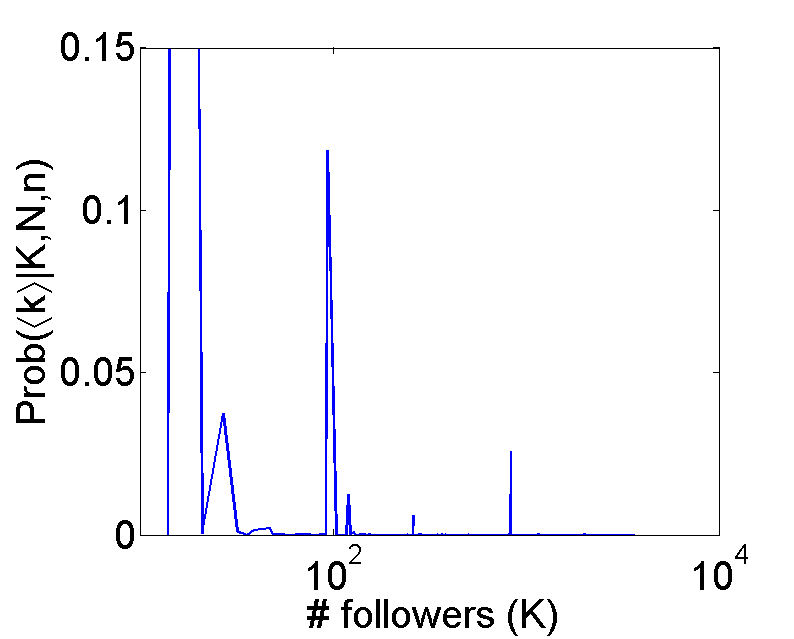}  \\
 (a) & (b)\\
\multicolumn{2}{c}{Twitter} \\
  \includegraphics[width=0.5\linewidth]{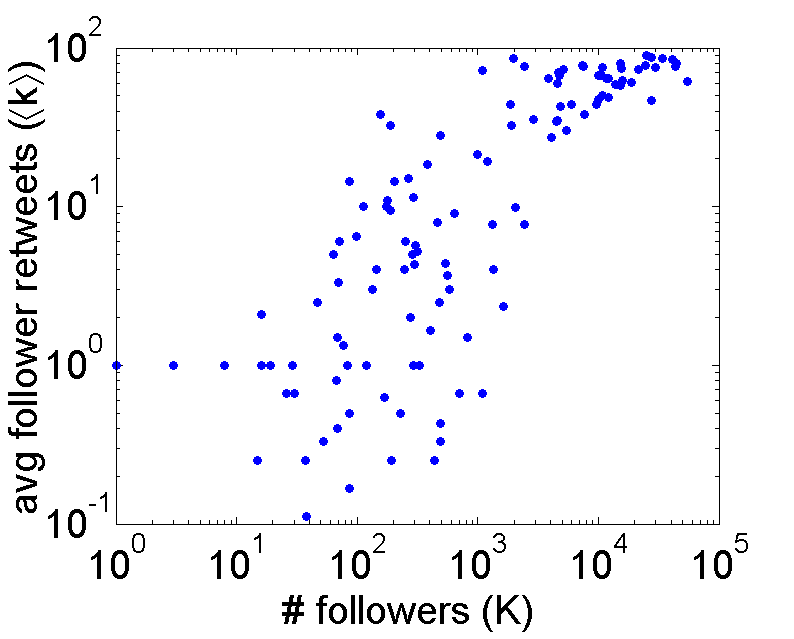} &
   \includegraphics[width=0.5\linewidth]{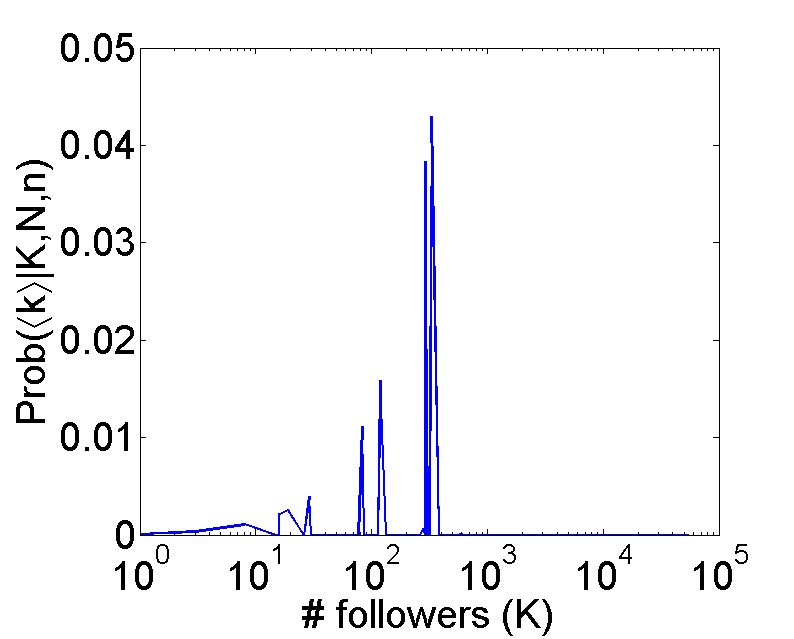}  \\
 (c) & (d)
\end{tabular}
\caption{Analysis of the empirical estimate of influence on Digg and Twitter. (a, c) The scatter plot shows the average number of times followers rebroadcast a story within its first 100 rebroadcasts vs. the number of followers the submitter has. Each point represents a distinct submitter. (b, d) Probability of the expected number of follower rebroadcasts being generated purely by chance. }\label{fig:avg_fan_votes}
\end{figure}

\subsubsection{Measuring local influence on Digg}
To reduce the effect of the front page to which Digg promotes popular stories, we count the number of votes from submitter's followers within the first 100 votes only. Since few stories are promoted to the front page before they receive that many votes, this ensures social links are mainly responsible for spreading interest in stories~\cite{Lerman10icwsm}.
Of the 3552 stories in the Digg data set,  3489 were submitted by 572 connected users.
Of these, 289 distinct users submitted two or more  stories which received at least one follower vote within the first 100 votes, providing us with enough information to estimate influence.
Figure \ref{fig:avg_fan_votes}(a) shows the average number of follower votes $\langle k \rangle$ within the first 100 votes received by stories submitted by these users  versus the number of followers $K$ these users have.

Are these observations significant? Do submitters with more followers simply get more votes due to greater numbers of followers? Or could we have observed that many follower votes purely by chance?
Let's assume that there are $N$ users who vote for stories randomly, independently of who submits them.
This type of stochastic voting can described by the \emph{urn model}~\cite{urn}.
Imagine an urn that contains $N$ balls, of which $K$ are white. Imagine also that we draw $n$ balls from the urn without replacing them. How many of them will be white? The probability that $k$ of the first $n$ votes come from submitter's followers purely by chance is equivalent to the probability that $k$ of the $n$ balls drawn from the urn are white.
This probability is given by the hypergeometric distribution:
 \begin{equation}
P(X=k|K,N,n)= \frac{ \left( \begin{array}{c}
K  \\
k  \end{array} \right)  \left( \begin{array}{c}
N-K  \\
n-k  \end{array} \right)}{\left( \begin{array}{c}
N  \\
n  \end{array} \right)}
\label{eq:hypergeometric}
\end{equation}

Using Eq.~\ref{eq:hypergeometric}, we compute the probability $P(X=\langle k \rangle|K,N,n)$ (N=71367, n=100) a story submitted by a Digg user with $K$ followers received $\langle k \rangle$ votes from submitter's followers purely by chance. As shown in Figure~\ref{fig:avg_fan_votes}(b), for $K>100$, this probability is very small; therefore, it is unlikely ($P<0.05$) these votes could arise purely by chance. We conclude that average number of follower votes received by stories submitted by a user (with at least 100 followers) is
a statistically significant ($P<0.05$) measure of her influence.

\subsubsection{Measuring local influence on Twitter}
We analyzed the Twitter data set using the same methodology. There were 174 users who posted at least two URLs that were retweeted at least 100 times.
Figure~\ref{fig:avg_fan_votes}(c) shows the average number of times the posts of these users were retweeted by their followers. Figure~\ref{fig:avg_fan_votes}(d) shows the probability these number of retweets could have been observed purely by chance. Since these values are small, we conclude that average number of follower retweets is a statistically significant ($P<0.05$) estimate of influence on Twitter.

\subsubsection{Measuring global influence}
Alternatively, we can measure the influence of the submitter by the average size of the cascades her posts trigger. A cascade describes how information spreads on the follower graph. The cascade begins with a seed, e.g., story submitter, who broadcasts the story to her followers. It grows when these followers choose to vote or retweet the story, in turn broadcasting it to their own followers, and so on.  All nodes in a cascade are connected to the seed through follower relations, either directly or indirectly though other nodes in the cascade.

For each post, we extracted the cascade that starts with the submitter and includes all voters/retweeters who are connected to voters/retweeters in the cascade via follower links. The larger the cascade size (on average), the more influential the submitter.

\subsection{Comparison of Centrality Measures}
 \begin{figure}[tbh]
\begin{tabular}{@{}c@{}c@{}}
\multicolumn{2}{c}{Digg} \\
  \includegraphics[width=0.5\linewidth]{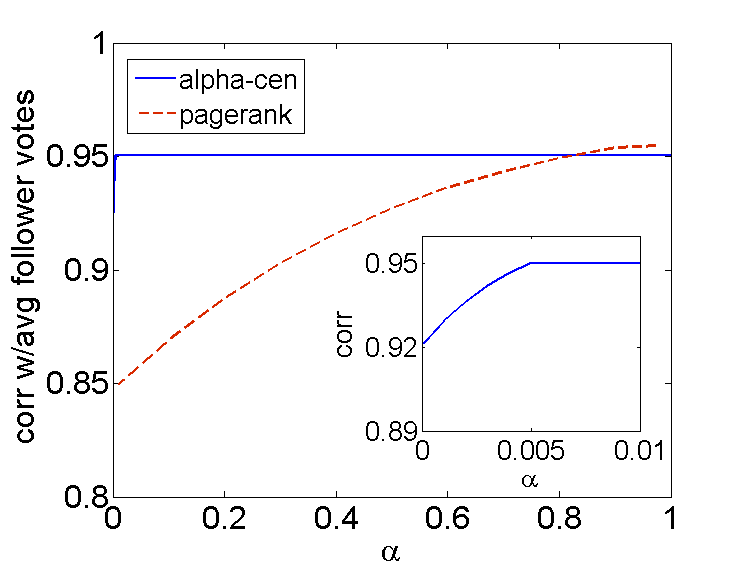} &
    \includegraphics[width=0.5\linewidth]{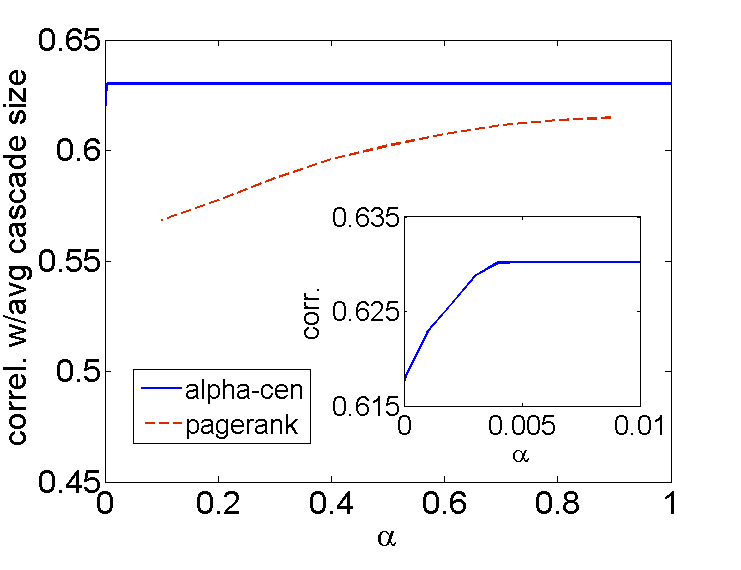} \\
   (a) Digg local & (b) Digg global\\
 \multicolumn{2}{c}{Twitter} \\
 \includegraphics[width=0.5\linewidth]{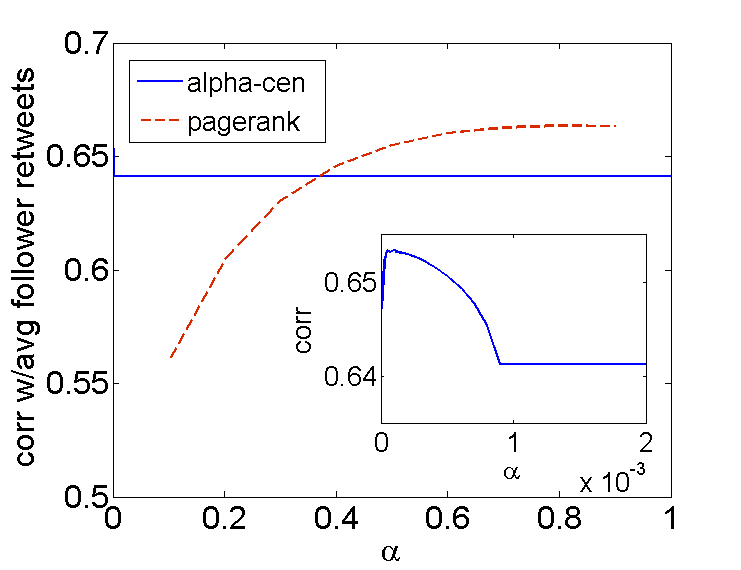} &
    \includegraphics[width=0.5\linewidth]{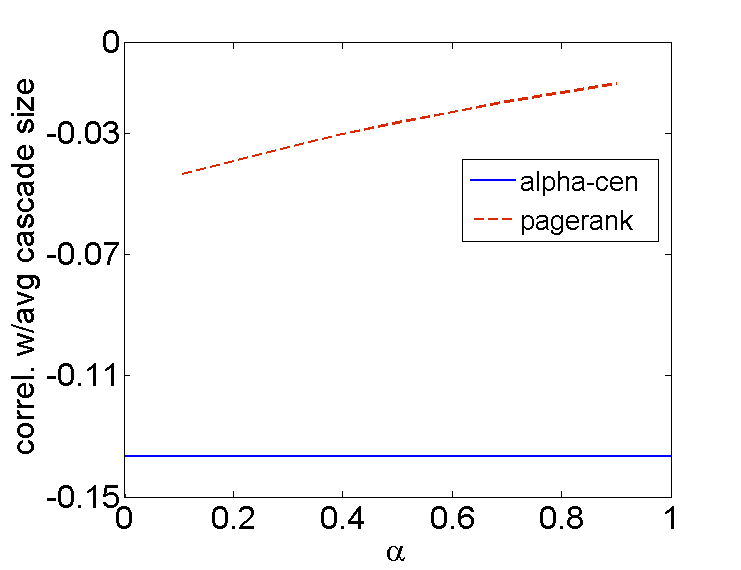} \\
 (c) Twitter local & (d) Twitter global
\end{tabular}
\caption{Correlation between the rankings produced by the local (average number of follower re-broadcasts) and global (average cascade size) empirical measures of influence with those predicted by normalized Alpha-Centrality and PageRank  on Digg and Twitter. }\label{fig:correl_digg}
\end{figure}

We use empirical estimates of influence to rank a subset of users in the Digg and Twitter data sets who submitted more than one story (URL) which received at least 100 votes (retweets).
However, of the 174 such submitters on Twitter, only 75 could be classified as not spammers according to the entropy criteria~\cite{Ghosh11snakdd} mentioned above; therefore, we restrict analysis to these users.
We evaluated centrality measures by comparing how they rank these submitters with how they are ranked by the empirical measures of influence using Pearson's correlation coefficient. We studied standard PageRank (with uniform starting vector) and Alpha-Centrality (with in-degree as the starting vector), both of which were computed on the follower graph. The effect of using other starting vectors for  PageRank (as is done in personalized PageRank)  and Alpha-Centrality is the course of future work.

Figure~\ref{fig:correl_digg} shows the correlation between the empirical measures of influence  with normalized Alpha-Centrality and PageRank on Digg (a,b) and Twitter (c,d). Parameter $\alpha$ stands for the \emph{attenuation factor}  for  Alpha-Centrality (see Equations~\ref{a-cen1} and \ref{n-a-cen-summation}) and the \emph{damping factor} (\emph{restart probability}=$1-\alpha$) for PageRank  (see Equation~\ref{eq:pr1}).
Figures~\ref{fig:correl_digg}(a) and \ref{fig:correl_digg}(c) show the correlation of the local measure of influence (average number of follower re-broadcasts), while Figures~\ref{fig:correl_digg}(b) and \ref{fig:correl_digg}(d) show the correlation of the global empirical estimate of influence (average cascade size).
On Digg (Fig.~\ref{fig:correl_digg}(a)), Alpha-Centrality correlates better than PageRank  with the local measure of influence over a wide range of $\alpha$ values; however, on Twitter, PageRank starts to perform better for $\alpha > 0.4$ (Fig.~\ref{fig:correl_digg}(c)). This could be because simple epidemics do not completely describe information spread in social media~\cite{Versteeg11,Hodas12}.

Though the correlation with global influence on Digg (Fig.~\ref{fig:correl_digg}(b)) is less overall than with local influence, Alpha-Centrality outperforms PageRank for all values of $\alpha$. Surprisingly, the correlations on Twitter (Fig.~\ref{fig:correl_digg}(d)) are negative. This is consistent observations of Bakshy et al.~\cite{Bakshy11}, who found that cascade size of past submissions was not a good predictor of the cascade size of a user's future submissions on Twitter. The correlations are less negative for PageRank, but it is difficult to conclude anything about the relative performance of Alpha-Centrality and PageRank from these results.

The insets show the interval corresponding to small values of $\alpha$. Note that normalized Alpha-Centrality becomes a global metric very quickly, i.e., over a small range of $\alpha$ values. The point at which it becomes constant corresponds to the epidemic threshold.
There are interesting differences in the behavior of correlation with the empirical measure of influence on Digg and Twitter. On Digg, the correlation with Alpha-Centrality grows from $\alpha=0$, suggesting that global structure becomes more important in determining influence, while on Twitter it has the opposite behavior. These differences could arise from differences in the network structure, and will be addressed in future research.

\remove{
\begin{figure}[tbh]
\begin{tabular}{c}
  \includegraphics[width=0.75\linewidth]{fig/digg_correlation_cascade_size_inset}  \\
   (a) Digg \\
  \includegraphics[width=0.75\linewidth]{fig/twitter_cascade_correlation} \\
 (b) Twitter
\end{tabular}
\caption{Correlation between the rankings produced by the global empirical measure of influence (average cascade size) and those predicted by normalized Alpha-Centrality and PageRank  on (a) Digg and (b) Twitter. }\label{fig:correl_cascade}
\end{figure}

Figure~\ref{fig:correl_cascade} shows the correlation of the global empirical estimate of influence (average cascade size) with Alpha-Centrality and PageRank for the same sets of submitters. Though the correlation with global influence on Digg is less overall than for local influence, Alpha-Centrality outperforms PageRank for all values of $\alpha$. Surprisingly, the correlations on Twitter are negative. This is consistent observations of Bakshy et al.~\cite{Bakshy11}, who found that cascade size of past submissions was not a good predictor of the cascade size of a user's future submissions on Twitter. The correlations are less negative for PageRank, but it is difficult to conclude anything about the relative performance of Alpha-Centrality and PageRank from these results.
}

The empirical results, for the most part, support our claim that Alpha-Centrality is better able to identify important users than PageRank because it more closely models the spread of information on social media, which takes place via broadcasts from users to their followers. However, PageRank sometimes outperforms Alpha-Centrality on the local measure of influence, indicating that information spread is a more complex process than a simple epidemic~\cite{Versteeg11,Hodas12}. Incidentally, $\alpha=0.85$, above which PageRank outperforms Alpha-Centrality on Digg, was the value suggested by Brin et al.~\cite{PageRank} for finding important pages in a Web graph. Empirical studies suggest different values of $\alpha$ are appropriate for different domains~\cite{Gleich10}, although some authors caution~\cite{Boldi05pagerankas} against using values of $\alpha$ close to one. Since it is not clear what value of $\alpha$ would be appropriate for social networks, Alpha-Centrality's better overall performance suggest that it is better suited for identifying influential users on Digg.

Results of correlation of centrality with the global measure of influence are less conclusive. While Alpha-Centrality does correlate better than PageRank with this measure on Digg for all values of $\alpha$, on Twitter these measures are anti-correlated.
One possible explanation could be differences in the user interface on these sites. Another possibility is that information spread deviates from a simple epidemic more on Twitter. Yet another explanation could be  differences in network structure on the two sites, or simply an artifact of the biases introduced by our aggressive spam filtering or small size of the data set. We are addressing these questions in our ongoing work.

\section{Related Work}
The interplay between structural properties of  networks and the diffusion processes occurring on them contribute to their complexity. This has been realized by several researchers in the past.
For example, Lambiotte et al.~\cite{Lambiotte09,Lambiotte10} emphasized that dynamical processes play an important role in characterizing the structure of complex networks.  In \cite{Lambiotte09} they measure the quality of a network partition in terms of the statistical property of the dynamic process taking place in the network. In \cite{Lambiotte10} they study the different equilibrium properties of these processes.
 However, their works focus on what we call conservative processes: unbiased and biased random walks, discrete and continuous time random walks. In contrast, we also study non-conservative dynamical processes.
We also relate these processes to centrality. Although the relationship of PageRank to random walk-type processes is well known, we explain how Alpha-Centrality is related to a type of a non-conservative process. We also carry out an empirical study of different centrality measures, unlike previous works.

Non-conservative processes are useful for studying a wide range of social phenomena, including the spread of epidemics within a population and information diffusion in social media, viral marketing, and many others. Many of these phenomena have been investigated by other researchers. The study of epidemics, in particular, has a very long history~\cite{Hethcote00,Bailey:1975}. It is known, for example, that epidemics exhibit critical behavior, and that the threshold of critical behavior is related to network structure~\cite{Wang03,Prakash11,Pastor-Satorras2001}.
The present work further confirms the relationship between epidemic threshold and network structure. Moreover, it gives an intuitive explanation for critical behavior of epidemics in terms of diverging length scales of non-conservative interactions.

Borgatti~\cite{Borgatti05} suggested a link between centrality and dynamical processes, defining a node's centrality in terms of its participation in the flow taking place on the network~\cite{Borgatti06}. Therefore, he claimed, the appropriate centrality measure for a given network is one that takes into account the details of the flow.
He proposed a typology of flows, based on the trajectories they follow (e.g., geodesics, paths, trails) and the mechanism of spread (e.g., transfer or broadcast),
and used simulations to explore the relationship between flows and centrality measures, such as betweenness, degree, and eigenvector~\cite{bonacich72factoring} centralities. He showed that centrality whose assumptions matched details of the flow was able to better reproduce key observations, such as how quickly or how frequently the flow reached a node. For example, a flow that follows geodesics (shortest paths) frequently visits nodes with highest betweenness centrality~\cite{Freeman79}.  We propose a simpler classification scheme that differentiates flows based on whether or not they conserve the flowing quantity. Unlike Borgatti's work, we mathematically explore the relationship between different flows and centrality and empirically study differences between centrality measures.

Estrada et al.~\cite{Estrada12}  studied measures similar to Alpha-Centrality and personalized PageRank (with attenuation factor 1) which they call communicability.  They linked the communicability functions to dynamics by showing their relationship to the thermal Green's function of oscillators. They used communicability to identify important actors in small social networks, demonstrating that different communicability functions led to different judgements of centrality, but did not justify the choice of the particular communicability function in terms of the interactions taking place between actors. Although we study a similar function, the goal of our work is to contrast conservative and non-conservative dynamics and explain how these differences should guide the choice of centrality measure for a given social network.

Researchers are increasingly turning to social media data sets to study the properties of complex networks. Some studies used activity-based measures, such as the number of mentions or re-tweets~\cite{Cha10icwsm,Bakshy11} to identify important social media users. Besides correlating these activity-based measures with degree centrality~\cite{Cha10icwsm}, no study has investigated centrality in social media. Our focus in this paper is to justify the choice of centrality by taking into account the dynamical processes taking place on the network.

\section{Conclusion}
We described two fundamentally distinct dynamical processes on networks, which can be differentiated based on whether or not they conserve some quantity that is distributed on the network,
and studied their relationship to two well-known centrality measures used to identify important or influential actors in a social network: PageRank and Alpha-Centrality.
While PageRank represents a steady state distribution of a conservative dynamic process on a network, for example, a random walk with restarts, we showed that Alpha-Centrality is a solution of non-conservative dynamics, examples of which include epidemics and signaling by broadcasts.

By analyzing data about information diffusion in social media, we found that Alpha-Centrality tends to better correlate with the empirical measures of influence than PageRank, although it is not clearly superior overall. Our recent research suggests that while information diffusion in social media does have a non-conservative flavor, it is a more complex process than a simple epidemic~\cite{Hodas14srep}. A centrality measure that takes into account the nature of information spread in social media could better predict influential social media users. We are currently studying the impact of the microscopic mechanics of contagion on centrality.

Centrality is but one type of measurement of network structure. Other types of measurements, for instance, community detection or determining the strength of social ties, may also be affected by the nature of the dynamic processes occurring on networks. We are addressing these in our ongoing work.

\section*{Appendix}
\label{ch:Appendix}
Replication matrix $\WN$ can be written  in terms of its eigenvalues and eigenvectors as:
\begin{equation}
\WN=X\Lambda X^{-1} = \sum_{i=1}^{|V|} \lambda_{i}Y_{i}
\label{eq:eigen}
\end{equation}
where $X$ is a matrix whose columns are the eigenvectors of $\WN$.
$\Lambda$ is a diagonal matrix, whose diagonal elements are the eigenvalues, $\Lambda_{ii}=\lambda_{i}$, arranged according to the ordering of the eigenvectors in $X$.
Without loss of generality we assume that $\lambda_{1} \ge \lambda_{2} \ge \cdots \ge \lambda_{n}$.
The matrices $Y_{i}$ can be determined from the product
\begin{equation}
Y_{i}=X {Z}_{i}X^{-1}
\label{eq:z}
\end{equation}
where $Z_{i}$ is the \emph{selection matrix} having zeros everywhere except for element ${(Z_i)}_{ii}=1$ ~\cite{Gebali:2008}.
Therefore,
\begin{eqnarray}
S({\alpha,t})&= &{\sum_{k=0}^{t}  (\alpha \WN)^k} \nonumber \\
&=& I+ \alpha \lambda_1{\displaystyle \sum_{i=1}^n} \frac{{(-1)}^{\mathcal{I}_i}\left(1-\alpha^{t+1} \lambda_{i}^{t+1}\right)}{{(-1)}^{\mathcal{I}_i}(1-\alpha \lambda_{i})} Y_{i}
\label{eq:cm}
\end{eqnarray}
where $\mathcal{I}_i=0$ if $\alpha \left |\lambda_{i}\right | <1$ and $\mathcal{I}_i=1$ if $\alpha \left |\lambda_{i}\right | >1$. As obvious from above, for  Equation \ref{eq:cm}  to hold non-trivially, $\alpha \neq \frac{1}{\left |\lambda_{i}\right|} \forall i \in 1,2\cdots,n$. Now assuming $|\lambda_{1}|$
    is strictly greater than any other eigenvalue
    $$S({\alpha,t}) \approx  I+ \frac{{(-1)}^{\mathcal{I}_1}(\alpha \lambda_1(1-\alpha^{t+1} \lambda_{1}^{t+1}))}{{(-1)}^{\mathcal{I}_1}(1-\alpha \lambda_{1})} Y_{1}.$$

  For any matrix $M$, let $||M||_1= \sum_{i,j} M[i,j]$  Therefore, the expected number of paths is $||S(\alpha,t)||_1$.
  The expected path length is given by:
\begin{eqnarray*}
 \frac{{\displaystyle \sum_{k=0}^t} k\alpha^{k}||\WN^{k}||_1}{{\displaystyle \sum_{k=0}^t} \alpha^{k}||\WN^{k}||_1} &=& \frac{\alpha \frac{d||S(\alpha,t)||_1}{d\alpha}}{||S(\alpha,t)||_1} \nonumber  \\
&\approx & {{(-1)}^{\mathcal{I}_1}  \left(\frac{1}{1-\alpha \lambda_1} - (t+1) \frac{\alpha^{t+1}\lambda_1^{t+1}}{1- \alpha^{t+1}\lambda^{t+1}}\right)}
\label{eq:cm1}
\end{eqnarray*}
Therefore,  as $t\to \infty$ and $\alpha|\lambda_1|<1$, the expected path length is approximately  $\frac{1}{1-\alpha \lambda_1}$, and for $\alpha|\lambda_1|>1$ it is $O(t)$.

\section*{Acknowledgments}
This material is based upon work supported by the National Science Foundation under Grants No. 0915678 and CIF-1217605,  the Air Force Office of Scientific Research under Contract Nos. FA9550-10-1-0102 and FA9550-10-1-0569, by the Air Force Research Laboratory under Contract No. FA8750-12-2-0186, and by DARPA under Contract No. W911NF-12-1-0034. KL would like to acknowledge Suradej Intagorn for collecting Digg data, Tawan Surachawala and Jeon-Hyung Kang for collecting and analyzing Twitter data. Authors would also like to thank Konstantin Voevodski for helpful comments and Shang-Hua Teng for the edifying and insightful conversations.

\providecommand{\href}[2]{#2}
\providecommand{\arxiv}[1]{\href{http://arxiv.org/abs/#1}{arXiv:#1}}
\providecommand{\url}[1]{\texttt{#1}}
\providecommand{\urlprefix}{URL }

\end{document}